\documentclass[twocolumn,prl,preprintnumbers,amsmath,amssymb]{revtex4}

\usepackage{graphicx}
\usepackage{dcolumn}
\usepackage{bm}
\usepackage{times}

\begin{document}


\title{Spin-selective optical absorption of singly charged excitons in a quantum dot}

\author{Alexander H\"{o}gele}
\author{Martin Kroner}
\author{Stefan Seidl}
\author{Khaled Karrai}
\affiliation{Center for NanoScience, Department f\"{u}r Physik, Ludwig-Maximilians-Universit\"{a}t,
Geschwister-Scholl-Platz 1, 80539 M\"{u}nchen, Germany}

\author{Mete Atat\"{u}re}
\author{Jan Dreiser}
\author{Atac Imamo\u{g}lu}
\affiliation{Institute of Quantum Electronics, ETH H\"{o}nggerberg HPT G12, CH-8093 Z\"{u}rich, Switzerland}

\author{Richard J. Warburton}
\affiliation{School of Engineering and Physical Sciences,
Heriot-Watt University, Edinburgh EH14 4AS, UK}

\author{Antonio Badolato}
\author{Brian D. Gerardot}
\author{Pierre M. Petroff}
\affiliation{Materials Department, University of California, Santa Barbara, California 93106, USA}

\date{October 14, 2004}

\begin{abstract}
We report high resolution laser absorption spectroscopy of a
single InGaAs/GaAs self-assembled quantum dot embedded in a
field-effect structure. We show experimentally that the interband
optical absorption to the lower Zeeman branch of the singly
charged exciton is strongly inhibited due to spin (Pauli) blockade
of the optical transition. At high magnetic fields the optical
absorption to the upper Zeeman branch dominates the absorption
spectrum. We find however that the spin blockade is not complete
and a 10\% leakage remains at high magnetic fields. Applying a
gate voltage to empty the dot of its resident electron turns the
spin blockade off. This effect is observed at 1.5~K and up to
9~Tesla.
\end{abstract}


\maketitle

The coherence time of an excess electron spin strongly confined in
a quantum dot (QD) structure is expected to be orders of magnitude
longer than the typical timescales required for its coherent
manipulation \cite{Golovach,Hanson,Kroutvar,Elzerman}. Motivated
by this observation, several groups have proposed to use single QD
spins as quantum bits (qubits) \cite{Loss}, and to manipulate,
couple and measure individual spins using either transport
\cite{Elzerman} or optical techniques \cite{Imamoglu}. In the case
of self-assembled InGaAs QDs with strong confinement along the
growth direction, the lowest energy optical transitions are those
arising from the excitation of a $J_{z}$=$+\frac{3}{2}$
$(J_{z}$=$-\frac{3}{2})$ valence electron to a
$S_{z}$=$+\frac{1}{2}$ $(S_{z}$=$-\frac{1}{2})$ conduction state.
If the QD already has an excess conduction electron, only one of
these optical transitions is allowed; the other is spin (Pauli)
blocked \cite{Warburton1}. In contrast, a neutral QD with
asymmetric confinement potential always has a pair of exciton
transitions. It has been suggested that Pauli blocking of
absorption or fluorescence can be used to implement high
efficiency all-optical single-spin measurements
\cite{Imamoglu,Calarco} and conditional spin dynamics
\cite{Calarco}.

In this Letter, we report resonant absorption measurements on a
single QD charged with a single excess electron in the regime of
Pauli blocking. We observe that for high magnetic fields
(B$>$5~Tesla) where the electron is (with high probability) in the
lowest energy $S_{z}$=$+\frac{1}{2}$ state, the absorption of a
left-hand-circularly polarized ($\sigma ^{-}$) laser is suppressed
by a factor of 10 as compared to that of a right-hand-circularly
polarized ($\sigma ^{+}$) laser. In contrast, we observe that both
$\sigma ^{+}$ and $\sigma ^{-}$ transmission dips have equal
strength for a neutral QD, irrespective of the applied magnetic
field. Using the fact that the strength of $\sigma ^{+}$ and
$\sigma ^{-}$ absorption in a charged QD is a measure of the spin
polarization of the resident electron, we were able to determine
the electron g-factor by fitting the magnetic field dependence of
the transmission data. Our results represent a first step towards
all-optical coherent spin measurement.

\begin{figure}[t]
\includegraphics[scale=0.9]{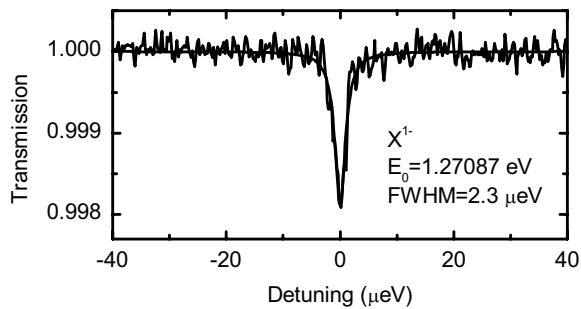}
\caption{\label{fig1}High resolution laser spectroscopy of the transmission
spectrum to a singly negatively charged exciton X$^{1-}$ confined
in a single self-assembled InGaAs quantum dot. The spectrum is
taken at 4~K and B=0~T. The data are fitted to a Lorentzian peaked
at E$_{0}$=1.27087~eV and with 2.3~$\mu$eV full width at half
maximum.}
\end{figure}

The InGaAs dots investigated in this work were self-assembled
in the Stranski-Krastanow growth mode by molecular beam
epitaxy. The QDs are separated from a highly n-doped GaAs back contact by 25~nm of intrinsic GaAs which acts as a tunnel barrier. The electrons are prevented from tunneling to the gate
electrode, the metalized sample surface, by a 110~nm thick
GaAs/AlAs superlattice blocking barrier. The whole structure forms
a field-effect device. The electron occupation of the dots is controlled by
applying a gate voltage and monitored with the QD photoluminescence charging diagrams
\cite{Warburton2}. An individual QD is identified to be spectrally
separated by typically more than 5~meV from neighboring dots assuring that the observed features originate
unambiguously from a single QD. We used high resolution laser
spectroscopy and a low temperature confocal microscope to measure
the differential transmission through the QDs. This technique \cite{Alen,PhysE,PRL} gives optical spectra as shown in
Fig.~\ref{fig1}. The spectral resolution is much less than the QD
absorption width. A dip in the transmission is obtained when the resonantly scattered laser light is diverted away from the photodetector placed immediately behind the sample.

In the absence of a magnetic field, a negatively charged exciton
(X$^{1-}$) strongly confined in a QD shows a single unpolarized
absorption peak in the optical interband spectrum, as demonstrated
in Fig.~\ref{fig1} with high resolution transmission spectroscopy.
In the case of X$^{1-}$, electrons are forced to fill the lowest
energy $s$-shell such that the total electron spin is 0 in
accordance with the Pauli principle. As a consequence, the
exchange interaction between the confined electrons and the hole
is suppressed so that the optical spectrum is free of fine
structure features irrespective of the asymmetry of the confining
potential \cite{PRL,Bayer1}. This is not the case for the neutral
exciton X$^{0}$: the electron-hole exchange leads generally to a
splitting into two linearly polarized dipole-allowed optical
transitions \cite{PRL,Gammon,Bayer2}. For the dot investigated
here the measured neutral exciton splitting was 10~$\mu$eV.

\begin{figure}[t]
\includegraphics[scale=0.9]{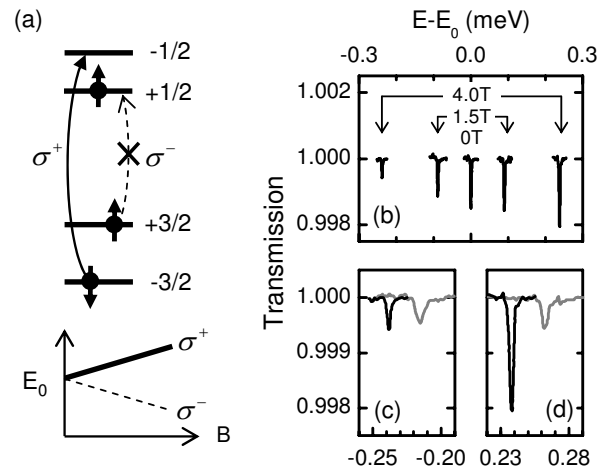}
\caption{\label{fig2}(a) Schematics of the quantum dot Zeeman-split highest
valence levels $J_{z}$ $(-\frac{3}{2},+\frac{3}{2})$ and the
lowest conduction levels $S_{z}$ $(+\frac{1}{2},-\frac{1}{2})$
labelled with their respective angular momentum quantum numbers.
The Zeeman energies are given by
$\mu_{B}B(g_{e}S_{z}-\frac{1}{3}g_{h}J_{z})$. Here the Land\'e
factors are $g_{e}<0$ and $g_{h}<0$. The dipole-allowed
transitions are shown with their circular polarizations $\sigma
^{+}$ and $\sigma ^{-}$. The low energy optical transition is
Pauli blocked by the presence of the resident electron in the
lowest conduction level. The schematic evolution of the absorption
energy in magnetic field B is also shown. (b) Differential
transmission spectra of the charged exciton in a single quantum
dot at 1.5~K and magnetic fields of 0, 1.5 and 4.0~T. (c), (d) Low
and high energy resonances of the charged exciton transition at
4.0~T for 1.5~K (black) and 12~K (gray). The small energy shift
between the spectra is due to the slight dependence of the band
gap on temperature. For all spectra the laser polarization was
linear.}
\end{figure}

In a high magnetic field and at low enough temperatures, the
ground state of a singly charged QD is that of a spin-polarized
electron in the lowest Zeeman level. As shown in Fig.~\ref{fig2}a,
an interband photon can only be absorbed under the condition that
the photo-generated electron is of opposite spin. In principle, at
T = 0~K this translates into a single absorption line in the
optical spectrum irrespective of the applied magnetic field. In
spite of the applied magnetic field, the exciton absorption should
not show a Zeeman splitting, i.e. the optical transition at the
lower exciton Zeeman energy should be spin blocked due to the
Pauli principle independently of the photon polarization.

\begin{figure}[t]
\includegraphics[scale=0.9]{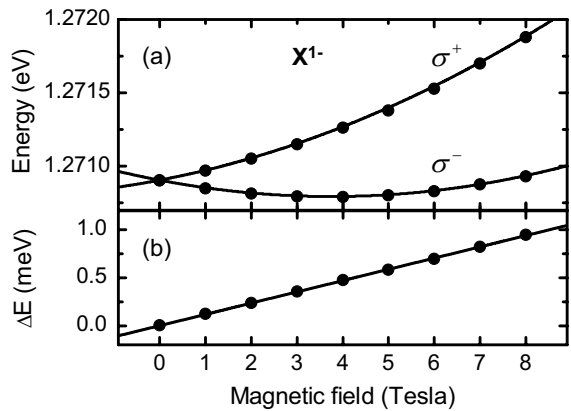}
\caption{\label{fig3}(a) Magnetic field dispersion of the charged exciton at
4.2~K. Both circularly polarized branches are shown. The solid
lines are fits to the data with $E=E_{0}\pm \frac{1}{2}
g^{*}\mu_{B}B+\beta B^{2}$ taking the exciton g-factor
$g^{*}=(g_{e}+g_{h})=-2.0$ and the diamagnetic shift
$\beta=7.9~\mu$eV/T$^{2}$. (b) Zeeman splitting of the two
branches. The solid line is a linear fit to the data with a slope
of 118~$\mu$eV/T.}
\end{figure}

\begin{figure}[b]
\includegraphics[scale=0.9]{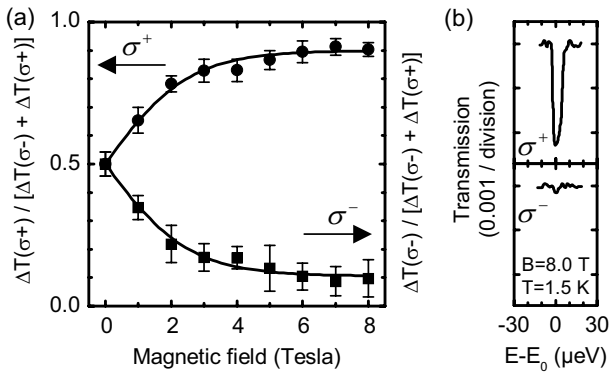}
\caption{\label{fig4}(a) Evolution of the relative resonance intensities with
magnetic field for both circular polarizations. The solid lines
are fits to the data with electron g-factor
$\mid$g$_{e}$$\mid$=1.8 as described in the text. The temperature
was 1.5~K. (b) Differential transmission spectra of the right-hand
circular (upper panel) and left-hand circular (lower panel)
transition at a magnetic field of 8.0~T.}
\end{figure}

We have investigated Pauli blocking experimentally.
Fig.~\ref{fig2}b shows transmission spectra in the presence of a
magnetic field. At T = 1.5~K and under magnetic field two
well-resolved transmission resonances are observed. Both lines
correspond to the two optical transitions shown in
Fig.~\ref{fig2}a. The energy positions of the resonances are
plotted in Fig.~\ref{fig3}a showing clearly a behavior quadratic
in field, consistent with the exciton diamagnetic shift
\cite{Schulhauser}. The energy splitting, plotted in
Fig.~\ref{fig3}b, is proportional to the applied magnetic field
consistent with a Zeeman splitting of the charged exciton. We have
conducted the transmission measurements as a function of the light
polarization. At finite magnetic fields the two dipole-allowed
transitions have right and left circular polarizations.
Fig.~\ref{fig2}b, c and d show the corresponding spectra using
linearly polarized light. At low temperatures, the strength of the
low energy resonance decreases with increasing magnetic field
while the opposite is true for the higher energy resonance.
Fig.~\ref{fig4}a shows the relative weight of the resonances as a
function of magnetic field. For each magnetic field the
polarization was optimized to maximize independently each of the
two resonances. At high magnetic field it can clearly be seen that
the absorption of the higher energy branch dominates. The low
energy branch is dramatically inhibited, saturating at high field
at a level of one tenth of the stronger peak, without ever
completely disappearing (Fig.~\ref{fig4}b). The high magnetic
field limit confirms roughly the expected Pauli blocking behavior
in that one resonance clearly prevails.

A further confirmation of the Pauli blockade picture is obtained
by increasing the temperature to T = 12~K. In this case, the
spectra obtained at B = 4~T show two Zeeman-split absorptions with
comparable peak strengths (Fig.s \ref{fig2}c and \ref{fig2}d). The
interpretation is that the resident electron is thermally
activated and occupies both spin states with about equal
probability. Lowering the temperature to 1.5~K clearly favors the
higher Zeeman branch confirming that the spin polarization of the
resident electron is the source of the Pauli blocking.

At T = 1.5~K the degree of spin polarization depends on the
magnetic field. We have used a two-level Boltzmann statistical
distribution for the spin state occupation. Considering that the
relative strength of both resonance peaks is a measure of the spin
polarization of the resident electron, we fitted the evolution of
the data in magnetic field, as shown  in Fig.~\ref{fig4}a. As a
fit parameter we used a Land\'e factor for the resident electron
of $\mid$g$_{e}$$\mid$=1.8$\pm$0.2. This value is about twice as
large as the one reported for InGaAs/GaAs self-assembled QDs in
Ref.~\cite{Bayer2}. However, several other dots emitting around 1.3~eV close to the center of the inhomogeneous QD energy distribution exhibit an electron g-factor of 0.6 as measured at 4.2~K. The deviation of the g-factor from dot to dot is not surprising since the value strongly depends on the individual In content and strain distribution. To obtain an optimal fit, we needed to
introduce empirically a limit to the maximum achievable electron
spin polarization of about 90\% in order to account for the fact
that the lower energy absorption branch does not vanish at high
fields, as shown in Fig.~\ref{fig4}b. This result indicates that
the Pauli blockade of the optical transition is not complete as
ideally predicted. It was also observed for the dots with the electron g-factor of 0.6 consistent with the temperature and the Zeeman energy. At the present time, the precise origin for the
apparent limitation of the Pauli blocking is still unknown and
requires more detailed measurements and analysis.

In conclusion, using high-resolution transmission spectroscopy we
have confirmed that a singly negatively charged exciton confined
in a single self-assembled QD has a dominant absorption line at
the higher Zeeman energy branch as a result of the spin
polarization of the resident electron. Unexpectedly, at high
fields the absorption at the lower Zeeman branch, although
drastically inhibited as expected from Pauli blocking, is still
not completely extinguished.

We would like to thank A. O. Govorov for helpful discussions.
Financial support for this work was provided in Germany by the DFG
grant no. SFB~631, in Switzerland by NCCR Quantum Photonics grant
no.~6993.1 and in the UK by the EPSRC.

\end{document}